\documentclass[12pt]{article}

\newcommand{\be}{\begin{equation}}
\newcommand{\ee}{\end{equation}}
\newcommand{\bea}{\begin{eqnarray}}
\newcommand{\eea}{\end{eqnarray}}
\newcommand{\RE}{{\rm Re}}
\newcommand{\im} {{\rm Im}}

\newcommand{\intL} {{\int_{4m_\pi^2}^{\infty}}}
\newcommand{\dint}[2]{\intL dz \frac{#1}{#2(z-s-i\epsilon)}}

\newcommand{\bd}{\begin{displaymath}}
\newcommand{\ed}{\end{displaymath}}

\newcommand{\epe}{\varepsilon'/\varepsilon}

\newcommand{\OM}{Muskheli\-sh\-vi\-li-Omn{\`e}s}

\begin{document}

%%%%%%%%%%%%%%%%%%%%%%%%%%%%%%%%%%%%%%%%%%%%%%%%%%%%%%%%%%%%%%%%%%%%%%%%%
% The titlepage
%%%%%%%%%%%%%%%%%%%%%%%%%%%%%%%%%%%%%%%%%%%%%%%%%%%%%%%%%%%%%%%%%%%%%%%%%

\author{
{\large\bf  A.J.~Buras${}^{1}$, M. Ciuchini${}^{1,2}$,
E. Franco${}^{1,3}$,}  \\
{\large\bf G. Isidori${}^{1,4}$, G. Martinelli${}^{3,5}$
 and L. Silvestrini${}^{1,3}$} 
}
\date{}
\title{
{\normalsize\sf
\rightline{TUM-HEP-366/00}
}
{\normalsize\sf
\rightline{LPT-Orsay/00-16}
}
\bigskip
{\LARGE\bf
Final State Interactions and $\epe:$\\ A Critical Look
}}

\maketitle
\thispagestyle{empty}

%\phantom{xxx}\vspace{-6mm} 

\begin{center}
{\small 
${}^{1}$ Physik Department, Technische Universit{\"a}t M{\"u}nchen,
 D-85748 Garching, Germany \\
${}^{2}$ Dipartimento di Fisica, Universit{\`a} di Roma Tre and INFN, 
 Sezione di Roma Tre, \\ Via della Vasca Navale 84, I-00146 Rome, Italy \\
${}^{3}$ Dipartimento di Fisica, Universit{\`a} di Roma ``La Sapienza''
 and INFN, Sezione di Roma, \\ P.le A. Moro 2, I-00185 Rome, Italy \\
${}^{4}$ INFN, Laboratori Nazionali di Frascati, 
 P.O. Box 13, I-00044 Frascati, Italy \\
${}^{5}$ LAL and Laboratoire de Physique Th{\'e}orique,  
 Laboratoire associ{\'e} CNRS, \\
 Universit{\'e} de Paris XI, B{\^a}t 211, 91405 Orsay Cedex, France
}
\end{center}

\bigskip

\begin{abstract}
We  critically analyze recent attempts to include final state interaction
(FSI) effects  in the calculation
of the CP violating ratio $\epe$. In particular 
the approach of using the \OM\ equation
for the decay amplitudes is examined. 
We find that the values of the dispersive correction 
factors are very sensitive to initial conditions which 
are, at present, out of control, and
demonstrate that the  claimed large 
enhancement of $\epe$ through FSI
is questionable. We propose, instead, a different implementation of this
approach which may be useful to lattice determinations of weak amplitudes.
\end{abstract}

\newpage
\setcounter{page}{1}
\setcounter{footnote}{0}

%%%%%%%%%%%%%%%%%%%%%%%%%%%%%%%%%%%%%%%%%%%%%%%%%%%%%%%%%%%%%%%%%%%%%%%%%%%%%%%
% The main part of the paper
%%%%%%%%%%%%%%%%%%%%%%%%%%%%%%%%%%%%%%%%%%%%%%%%%%%%%%%%%%%%%%%%%%%%%%%%%%%%%%%
\section{Introduction}
\setcounter{equation}{0}

One of the crucial  issues for our understanding of CP violation is the
question whether the size of the observed
direct CP violation in $K_L\to\pi\pi$ decays, expressed in terms
of   the ratio $\epe$, can be described within the Standard Model. 
Experimentally the grand average,
 including the results from  NA31 \cite{barr:93}, 
E731 \cite{gibbons:93}, KTeV \cite{KTEV} and NA48 \cite{NA48} reads
\be
\RE(\epe) = (21.2\pm 4.6)\times 10^{-4}\, .
\label{ga}
\ee

There are different opinions whether this result can be accommodated 
within the Standard Model. In \cite{EP99,ROMA99} the central value
of the predicted $\epe$ is typically
by a factor of 2-3 below the data, with an estimated theoretical
uncertainty of the order of $60 \div 100$~\%.  
The result in Eq.~(\ref{ga}), then,
can only be accommodated if all relevant parameters are chosen
simultaneously close to their extreme values. Higher values of $\epe$,
compatible with the data within one standard deviation have
been found, instead,  in \cite{BERT98,Dortmund}. Recent reviews can be found 
in \cite{BUJA}.

More recently,  following the previous work of Truong~\cite{truong},
it has been pointed out by Pallante and Pich~\cite{pp} 
that the inclusion of final state interactions substantially 
affects the estimates of $\epe$. The size of the effect is parametrized in
terms of dispersive correction  factors ${\cal R}_I$ 
multiplying the corresponding isospin amplitudes $A_I$. With 
${\cal R}_0\approx 1.4$ and ${\cal R}_2\approx 0.9$, as found in their numerical
evaluation,   a substantial enhancement of $\epe$ was  obtained. Indeed, 
by including these factors, the authors of Ref.~\cite{pp}
 find that the central value $\epe=7.0\times  10^{-4}$
of \cite{EP99} gets increased to $\sim 15\times 10^{-4}$ which is
much closer  to the experimental average (\ref{ga}). Similar results, although
with a different dispersive approach,  have been found in~\cite{PA99}.

The interesting paper of Pallante and Pich stimulated us to give a closer look
to the whole procedure followed to obtain the correction factors ${\cal R}_I$.
The phenomenological application of the \OM\ equation to the $K\to\pi\pi$ case
requires some arguable assumptions (see Section~\ref{sectf}).
However, even within those assumptions, we found a major objection which
unfortunately makes the calculation of Ref.~\cite{pp} questionable. On the
basis of symmetry arguments only, one can show that there is a point, out of
the physical cut, where the amplitudes induced by left-handed operators are
expected to vanish. As demonstrated below, the authors of Ref.~\cite{pp}
made an assumption on the
value of the derivative of the amplitude at this point
which is not justified.
All their results rely on this assumption.
We will show that, by replacing the latter condition 
with others, equally acceptable on physical grounds,
it is possible to obtain very
different values of the correction factors ${\cal R}_I$.
Thus  we conclude that the results found in \cite{pp} are
subject to substantial uncertainties. 

In spite of this, we found some interesting application of the dispersive 
analysis to non-perturbative  calculations of weak amplitudes. For example,
lattice calculations could provide a condition which replaces the arbitrary
assumption on the derivative of the amplitude
and fix unambiguously the factors ${\cal R}_I$ 
(under the hypotheses inherent to the use of the \OM\ equation
in $K\to\pi\pi$ decays).

Our paper is organized as follows. In section 2  we present general
formulae for the \OM\ solution of dispersion relations applied
to $K\to\pi\pi$ amplitudes, focusing on the assumptions 
needed to obtain the final result.
Using these formulae we demonstrate in Section 3 that the evaluation
of the dispersive factors ${\cal R }_I$ along the lines proposed in \cite{pp}
is ambiguous. In particular we stress that there is no justification
to use the numerical values of ${\cal R}_I$ found in \cite{pp} in 
conjunction with present lattice and large-$N_C$ 
calculations of hadronic matrix
elements. Similar comments apply to \cite{PA99}. 
In Section 4 we briefly illustrate how modifications of the technique
developed in \cite{truong,pp} could in principle improve the
accuracy of the predictions for $\epe$ in conjunction
with future lattice calculations. We conclude in Section 5.

\section{General formulae}
In this section we recall the basic ingredients used to derive the solution
of the \OM\ equation~\cite{omnes} in the case of  $K\to\pi\pi$ amplitudes.
The starting point is the $N$-subtracted dispersion relation. We denote
by $A(s)$ the decay amplitude of a kaon of mass $m_K=\sqrt{s}$ into
two pions of mass $m_\pi$. The invariant mass of the two pions is taken as
$(p_1+p_2)^2=s$, corresponding to the insertion of a weak operator 
carrying zero momentum.
The following discussion applies as well to the case where $A(s)$ is the
$K\to\pi\pi$ matrix element of a local operator renormalized at a 
fixed scale $\mu$.

Following Ref.~\cite{truong} we assume that $A(s)$ is analytic in the cut $s$
plane, with the cut going from $4m_\pi^2$ to
$\infty$~\footnote{~These analiticity properties of $A(s)$ are certainly not
fully correct. However one can argue that Eq.~(\ref{eq:disprel})
still holds to a reasonable accuracy in a region close to the physical point
(see Section~4).}. Further assuming that $s^{-N} A(s)\to 0$ for $s \to
\infty$, we can write \be A(s)=P_{N-1}(s,s_0)+\frac{(s-s_0)^N}{\pi}\dint{\im
A(z)}{(z-s_0)^N} \, , \label{eq:disprel}
\ee
where $P_{N-1}(s,s_0)$ is a $N-1$ degree $s$ polynomial, the coefficients
of which are to be fixed by $N$ independent conditions on the amplitude
and $s_0$ denotes the subtraction point (in general one could
choose $N$ different subtraction points), which has to lie outside
the physical cut. Writing the imaginary part of the amplitude as
\be
\im A(s)=A(s)e^{-i\delta(s)}\sin\delta(s),
\label{eq:eluni}
\ee
the dispersion relation (\ref{eq:disprel}) becomes a 
\OM\ equation~\cite{omnes}. Here 
$\delta(s)$ is the strong phase of the amplitude, which in the elastic region
is equal to the phase shift of the $\pi\pi$ scattering appropriate to
a given isospin channel~\footnote{~Note that even in the elastic region
$\delta(s)$ can be identified with the
phase shift measured from $\pi\pi$ scattering only if the dependence
of the latter on the kaon mass is negligible. We thank G.~Colangelo for
pointing this out to us.}. The
general solution of this equation is given by \cite{omnes}
\be
A(s)=Q_{N-1}(s)\, O(s)\,,
\label{eq:omnes}
\ee
where
\be
O(s)=\exp\Bigg(\frac{1}{\pi}\dint{\delta(z)}{}\Bigg)
\label{eq:omnesfact}
\ee
and $Q_{N-1}(s)$ is a $N-1$ degree
polynomial. Equation (\ref{eq:omnes}) shows explicitly 
that $N$ independent conditions on $A(s)$ 
and the knowledge of the phase on the whole cut are
sufficient to fully determine the amplitude.
Note that there is no longer any reference to the subtraction point
$s_0$ of the original dispersion relation.

We  point out that, if we were able to obtain 
a number of conditions on $A(s)$ larger than $N$, 
we could reduce the sensitivity of the solution 
to the knowledge of $\delta(s)$ at large $s$.
Indeed,
we can always rewrite the solution (\ref{eq:omnes}) as
\bea
A(s) &=& Q_{N-1}(s) \exp\Bigg(\sum_{i=0}^{M} c_i s^i \Bigg)
\nonumber \\ 
&& \times \exp\Bigg(\frac{s^{M+1}}{\pi} \intL dz
\frac{\delta(z)}{z^{M+1}(z-s-i\epsilon)} \Bigg)\, ,
\label{eq:marti}
\eea
where the product of the first two terms is completely 
determined by $N+M$ independent conditions.
If possible, this could be   useful since, in practice, $\delta(s)$ can be
extracted from the data only up to the inelastic threshold.

To make contact with the work of Pallante and Pich~\cite{pp}, 
we now discuss two examples  on how $A(s)$ is determined from a specific set of 
conditions  in the case $N=2$.  In this case,
corresponding to an amplitude going at most linearly in $s$ for large $s$,
two conditions on the amplitude are required:
\begin{enumerate}
\item In order to fix the coefficients of
$Q_1(s)$, a possible set of conditions is given by the knowledge of the
amplitudes at two different points $s_1$ and $s_2$, namely
\be
A(s_1)=A_1,\qquad A(s_2)=A_2\,.
\label{eq:inico}
\ee
In this case, one finds
\be
A(s)=\frac{A_1\,O(s_2)\,(s-s_2)+A_2\,O(s_1)\,(s_1-s)}
{(s_1-s_2)O(s _1) O(s_2)}\, O(s) \, . \label{eq:om1}
\ee
We stress that, although the
r.h.s of this equation is written in terms of $s_{1,2}$,
the physical amplitude is obviously independent of the choice
of these points, provided that $A_{1,2}$ are given by
Eq.~(\ref{eq:inico}).
\item A different set of conditions can be given by  
\be
A(s_1)=0\,,\qquad A^\prime(s_1)=C\,,
\label{eq:om2b}
\ee
namely by knowing the point $s_1$ where the amplitude vanishes, or
assumes a specific value, and its first derivative in that point.
With the conditions in Eq.~(\ref{eq:om2b})
the solution is particularly simple and takes the form
\be
A(s)=C(s-s_1)\,O(s_1)^{-1}\,O(s)\,.
\label{eq:om2}
\ee
\end{enumerate}

\section{Why we are not able to compute universal dispersive factors}
We now critically examine the procedure followed 
by Pallante and Pich  \cite{pp}
to compute the so-called {\it dispersive factors} 
${\cal R}_I$ for the $I=0$ and $I=2$ amplitudes.
To this end, we start by summarizing the main steps of 
the procedure adopted in Ref.~\cite{pp}.

On the basis of the lowest-order chiral Lagrangian,
Pallante and Pich assume that 
the solution of the \OM\ equation is given by
\be
A_I(s)=C_I(s-m_\pi^2)\,\Omega_I(s)\, ,
\label{eq:ppsol}
\ee
where
\be
\Omega_I(s)=\frac{O_I(s)}{O_I(m_\pi^2)}\equiv e^{i\delta_I(s)}{\cal R}_I(s)\, .
\ee
Here $O_I$ is given by
Eq.~(\ref{eq:omnesfact}) with $\delta$ replaced by $\delta_I$, the
experimental elastic $\pi\pi(I)$ phase shift, integrated up to a certain
cutoff. This solution corresponds to the case in Eq.~(\ref{eq:om2})
with $s_1=m_\pi^2$, i.e. the authors of Ref.~\cite{pp}
impose $A_I(m_\pi^2 )=0$
and  implicitly assume the knowledge of  
$A_I^\prime(m_\pi^2 )=C_I$.\footnote{~The need of these two
assumptions is common also to Ref.~\cite{truong}, as 
pointed out in \protect\cite{stech}.} 
In order to fulfill the latter condition, they assume without
justification that the matrix element at the physical point,
\be
{\cal M}_I=\langle\pi\pi(I)\vert{\cal H}_{\rm eff}\vert K\rangle
\big\vert_{s=m_K^2}\, , 
\ee
provided by non-perturbative methods such as lattice QCD or large $N_C$
estimates, corresponds  to $C_I(s-m_\pi^2)\vert_{s=m_K^2}$.
Their solution can thus be written as 
\be
A_I(s)={\cal M}_I\frac{(s-m_\pi^2)}{(m_K^2-m_\pi^2)}
e^{i\delta_I(s)}{\cal R}_I(s)
\label{eq:ppsol2}
\ee
and the dispersive factors ${\cal R}_I$ are simply given by
\be
{\cal R}_I\equiv{\cal
R}_I(m_K^2)=\left\vert \frac{A_I(m_K^2)}{{\cal M}_I}
\right\vert=\left\vert\frac{O_I(m_K^2)}{O_I(m_\pi^2)}\right\vert\,
. \ee
We have the following comments on this approach:
\begin{itemize}
\item One of the two conditions used to determine the solution
is $A_I(m_\pi^2)=0$. This is justified in Ref.~\cite{pp}
on the basis of
the lowest-order chiral realization of $(8_L,1_R)$ and $(27_L,1_R)$ operators.
The zero at $s=m_\pi^2$  of the amplitudes induced by these operators actually
holds beyond the lowest order \cite{KMW,BPP}. 
Indeed, it is a consequence of the vanishing of these amplitudes
in the $SU(3)$ limit, as follows by an old result by Cabibbo and
Gell-Mann~\cite{cgm}. In modern language, we can rephrase the
result of Ref.~\cite{cgm} by saying that the on-shell $K\to 2 \pi$ 
matrix element of any local $(8_L,1_R)$ operator invariant under
$CPS$ symmetry \cite{CPS} vanishes in the absence of $SU(3)$ breaking.
Therefore we think that this condition is well justified
in the case of $(8_L,1_R)$ operators and, particularly, in the case of
$Q_6$ considered in \cite{pp}.

\item 
Our main criticism is related to the determination of the second
condition on $A_I(s)$, namely the one on $A_I^\prime(m_\pi^2)$.
Chiral perturbation theory alone cannot fix the value of this 
derivative, which must be known from some non-perturbative information.
In their procedure, Pallante and Pich implicitly assume that the matrix
elements determined by non-perturbative methods such as lattice or
large $N_C$, fix the value of $A_I^\prime(m_\pi^2)$ according to the
condition \be
A_I^\prime(s=m_\pi^2)=C_I=\frac{{\cal M}_I}{m_K^2-m_\pi^2}\,.
\ee
This assumption, which may look
reasonable on the basis of the lowest-order chiral Lagrangian, actually
involves an ambiguity of the same order of the dispersive correction
itself. To illustrate this point,
we make a different choice of the initial conditions (while keeping the
same ${\cal M}_I$):
we use ${\cal M}_I$ to fix the value of the amplitude at threshold, via the
relation \be
A_I(s=4m_\pi^2)={\cal M}_I\frac{3m_\pi^2}{m_K^2-m_\pi^2}\, .
\ee
In other words, we assume that 
${\cal M}_I(s-m_\pi^2)/(m_K^2-m_\pi^2)$ provides a good approximation 
to the real amplitude near $s=4 m_\pi^2$ instead of the point $s=m_\pi^2$ 
as was implicitly employed in~\cite{pp}.
One may argue that this is also a reasonable choice because
strong-interaction phases  
vanish at threshold, but of course this condition is as arbitrary as the one
adopted in \cite{pp}.
Using this condition and $A(s=m_\pi^2)=0$, from
Eq.~(\ref{eq:om1}) we find
\be
A_I(s)={\cal M}_I\frac{(s-m_\pi^2)O_I(s)}{(m_K^2-m_\pi^2)O_I(4m_\pi^2)}\,.
\label{eq:oc} \ee
With the same
parameterization of the phase $\delta_0(s)$ as in Ref.~\cite{pp} and  our choice
(\ref{eq:oc}),
the dispersive factor for $(8_L,1_R)$ operators is 
\be
{\cal R}_0^{\rm threshold}=\left\vert \frac{O_0(m_K^2)}{O_0(4 m_\pi^2)}
\right\vert \simeq 1.1
\ee
instead of
\be
{\cal R}^{\rm PP}_0=\left\vert \frac{O_0(m_K^2)}{O_0(m_\pi^2)}
\right\vert \simeq 1.4\, .
\ee
With ${\cal R}_0=1.1$ (and ${\cal R}_2=1.0$) the central value 
$\epe=7.0\times  10^{-4}$ of \cite{EP99} gets increased 
to $\sim 9\times 10^{-4}$, instead of $\sim 15\times 10^{-4}$
found in \cite{pp}. While the first enhancement is well within
the theoretical uncertainties quoted by the various
analyses~\cite{EP99}--\cite{Dortmund}, the second one would have a
considerable impact on the predicted value of $\epe$.

Other choices of the initial conditions, equally acceptable under similar
assumptions, would lead to still different results for ${\cal R}_0$.
This exercise illustrates that, {\em unless one knows
the value of $s$} at which a given non-perturbative method provides the
correct value of the amplitude (and/or its derivative),
it is impossible to unambiguously compute the appropriate
correction factor due to final state interactions.

\item 
In the case of $\Delta I=3/2$ matrix elements of 
$(8_L,8_R)$ operators, such as $\langle Q_8^{(3/2)}\rangle$ considered
in Ref.~\cite{pp}, the amplitude does not vanish at $s=m_\pi^2$. Still 
the authors of \cite{pp} claim that the correction factor is ${\cal
R}_2=\vert O_2(m_K^2)/O_2(m_\pi^2)\vert \simeq 0.9$.
We can reproduce this result assuming that the dispersion relation requires
only one subtraction and that the non-perturbative calculations give the
correct amplitude at $s=m_\pi^2$. This condition, however, 
appears even more arbitrary than the one used to determine $A^\prime(m_\pi^2)$
in the case of $(8_L,1_R)$ operators. Indeed in this case 
the point $s=m_\pi^2$ does not play any special r{\^o}le.
It is therefore not surprising that in a different but consistent
framework, as the one discussed in \cite{DG,CG}, non-perturbative
corrections to $\langle Q_8^{(3/2)}\rangle$ are found to 
enhance, rather than suppress, lattice and large-$N_C$ estimates.
\end{itemize}
Let us then summarize the main point we want to make. While the full
amplitude clearly does not depend on the choice of the value
of $s$ at which the conditions on the amplitude are placed,
the dispersive correction factors are obviously dependent on
this choice as we explicitly demonstrated above.
As presently we are not in a position to state precisely which value
of $s$ the exisiting non-perturbative approaches, like
lattice, $1/N$, etc., correspond to,
it is not possible to determine uniquely the corresponding factors
${\cal R}_I$ to be incorporated in each of those methods.

Before concluding this section, we comment on other attempts 
present in literature to include FSI effects in  $\epe$.
The approach of Ref.~\cite{PA99} is somehow similar to 
the one of Ref.~\cite{pp}. In this paper, however, the 
integral over the cut is divided in two parts: a low-energy
region from threshold up to $\mu^2$ and a high-energy region 
from $\mu^2$ up to $\infty$.
By identifying the scale $\mu$ with the renormalization 
scale of the inserted operator, the high-energy region 
is claimed to be estimated in perturbation theory. 
The results thus obtained are in qualitative agreement 
with those of Ref.~\cite{pp} and show a rough independence from 
the unphysical scale $\mu$. 
We stress, however, that the matching procedure adopted in 
this paper cannot be justified theoretically.
Since we do not see a way of deriving consistently 
the formulae of Ref.~\cite{PA99} from first principles, 
we are  skeptical  about the approach followed in this paper.

Other recipes to include {\em a posteriori} FSI effects in non-perturbative 
calculations of weak amplitudes were presented in the recent 
literature~\cite{BERT98,Nierste}. For these we have objections
similar to those made to \cite{pp}: it is not clear to us how to relate the
theoretical values of the amplitudes, computed with some non-perturbative
method, to the physical ones which include FSI. In cases where the results
of non-perturbative calculations are unable to reproduce the experimental
phases, it is not possible to extract unambiguously the absolute value of the
amplitudes. Since corrections may be applied only if this issue is under
control, we think these procedures to incorporate empirically correcting factors
to account for FSI do not help to improve the accuracy of the calculation.
These attempts provide at best estimates of the theoretical
uncertainties, as for instance discussed in \cite{Dortmund}.

\section{Further developments}
\label{sectf}
In spite of the criticism discussed before, we believe the 
Truong-Pallante-Pich
approach is  interesting and deserves further investigation.
For this reason, in this section we briefly discuss possible
applications of this technique to improve the accuracy of
theoretical predictions for $K\to\pi\pi$ matrix elements.
We first want to state the assumptions needed to make this approach
useful:
\begin{enumerate}
\item 
In general, in order to fix the number $N$ of subtractions in the dispersion
relation, hence the degree of the polynomial $Q_{N-1}$, one has to make
an ansatz on the behaviour of the amplitude as $s\to\infty$. 
In the present case, however, Eq.~(\ref{eq:disprel}) should be understood
as an approximate equation, valid in a limited $s$ region close to the
physical point. Indeed other sources of non-analiticity besides the
cut starting at $4 m_\pi^2$ can appear in $A(s)$ due to the 
identification of $s$ with kaon mass, that is a parameter of the 
underlying theory. Still the solution of
Eq.~(\ref{eq:disprel}) can give a good estimate of the true amplitude if these
further sources of non-analiticity are well represented by a polynomial
expansion in the considered region. This is for instance the case in
an effective theory where the pions are the only dynamical fields, or
in the quenched approximation of QCD. In this context the number of
subtractions is no longer related to the behaviour of the amplitude
at large $s$, rather to the accuracy one needs to achieve.
Large values of $N$ are obviously preferable, although in
practice they are hardly attainable. Therefore, following
Refs.~\cite{truong,pp}, we assume that $N=2$ 
provides a reasonable approximation.
\item 
Above the inelastic threshold, $\delta(s)$ is no longer the $\pi\pi$
phase shift and cannot be directly extracted from the experimental data.
In addition, even in the elastic region, the phase of $A(s)$ can be
identified with the experimental $\pi\pi$  phase shift only assuming 
that the dependence of the $\pi\pi$ scattering amplitude 
on the kaon mass is negligible.
Were we able
to compute the weak amplitude for a large number of points, we could reduce the
sensitivity of the solution to the value of $\delta(s)$ at large
$s$, as shown in Eq.~(\ref{eq:marti}). As already
stressed, this possibility is presently rather remote. In practice, 
the integral entering $O(s)$ has to be restricted to the elastic region, unless
some further assumptions on the amplitudes of the other channels are made.
These issues entail some uncertainties which have been partially discussed in
Ref.~\cite{pp}.
\end{enumerate}

Under these assumptions, we now envisage a possible implementation of
this approach in lattice simulations. It has been shown that, even in
the presence of chiral symmetry breaking and of the Maiani-Testa
theorem~\cite{mt}, lattice calculations can obtain the physical matrix
elements at threshold, corresponding to
$s=4m_\pi^2$~\cite{dawson}. This
information, once available, combined with the constraint 
$A(s=m_\pi^2)=0$ and with the ansatz
that the amplitude grows linearly in $s$, is sufficient 
to determine the 
physical result for the real kaon mass (see Eq.~(\ref{eq:om1})) 
in the case of $(8_L,1_R)$ operators. 

Concerning $\Delta I=3/2$ matrix elements of 
$(8_L,8_R)$ operators, one may argue that one condition is 
not enough to fully determine the amplitude. We stress, however, 
that in this case the extrapolation between the $\pi\pi$ 
threshold and the physical kaon mass is expected to be
smoother, resulting in a correction factor which 
can be approximated by unity with high accuracy.

The evaluation of the amplitude at $s> 4m_\pi^2$ is particularly difficult
with lattice simulations since these are performed in 
the Euclidean space-time. Proposals,
however, exist for overcoming this problem~\cite{ciuchini,luscher}.
We finally stress
that the region $s< 4m_\pi^2$ is not accessible to lattice
calculations, as well as to any other approach, if the particles ought to be
on-shell and the inserted operator carries zero momentum.

\section{Conclusions}
We have critically discussed the issue of final
state interactions effects in the evaluation of the ratio 
$\epe$. Whereas we cannot exclude possible substantial 
enhancements of $\epe$ through FSI over lattice and
large-$N_C$ estimates, our analysis demonstrates
that the present calculation of the 
dispersive factors ${\cal R }_I$ is subject 
to considerable uncertainties, which prevent to draw 
any definite conclusion on the importance of these effects.

We summarize the main point of this paper. While the full
amplitude clearly does not depend on the choice of the value
of $s$ at which the conditions on the amplitude are placed,
the dispersive correction factors are obviously dependent on
this choice.
As presently we are not in a position to state precisely
which value of $s$ the exisiting non-perturbative calculations
correspond to,
it is not possible to determine uniquely the corresponding factors
${\cal R}_I$ to be incorporated in each of those methods.

On the other hand, we have shown that the \OM\ 
approach could be useful to reduce, or at least to better estimate, the
uncertainties on the physical amplitudes, once lattice results on
the matrix elements at threshold will be available.

\section*{Acknowledgments}
We are indebted to G.~Colangelo for critical comments and discussions about
the validity of dispersion relations in this context.
We also thank E.~Pallante and A.~Pich for correspondence.\\
This work has been supported in part by the German
Bundesministerium f{\"u}r Bildung und Forschung
under contract 06TM874 and 05HT9WOA0. 
E.F. and G.I. acknowledge a partial support
by the TMR Network under the EEC contract 
ERBFMRX-CT980169 (EURODA$\Phi$NE).
G.M. thanks the T31 group at the Technische
Universit{\"a}t M{\"u}nchen where part of this work
has been done. L.S. acknowledges a partial support by the M.U.R.S.T.

\end{document}